# Comment on *"The Analysis of EPR Spectra Using Tesseral Tensor Angular Momentum Operators"* by H. A. Buckmaster and R. Chatterjee

Czeslaw Rudowicz

*Department of Physics and Materials Science, City University of Hong Kong, Kowloon Tong, Hong Kong SAR*

Clear definitions of the operator notations and crucial notions used in the area of electron magnetic resonance (EMR) are important in order to avoid misunderstanding and incorrect results. For the benefit of a wider audience, in this Comment we attempt to clarify the confusing points, which exist in the paper in question.

## 1. Introduction

The paper [1] raises serious questions about the efficiency of the refereeing process to prevent publication of papers containing some faults and/or misconceptions. This may be partially due to the lack of collaboration between the Editorial Offices of various journals. The original manuscript [2] was rejected by another European physics journal in December 1995 with the present author acting as a referee. However, three years later it was published [1] with some modifications (see below) although no new theoretical or experimental results were presented therein. Moreover, the authors revisit, and sadly enough in the original manuscript [2] stick to, the faulty idea concerning the admission of the odd-order zero-field-splitting (ZFS) terms into spin-Hamiltonian (SH), put forward by them in earlier papers [3, 4]. The present author has discussed the question of the odd-order ZFS terms in SH with Prof. Buckmaster in person on two occasions and pointed out the refutation of this idea published in [5] (and also mentioned in [6]). Yet the authors [1] make no reference to [5], which not only perpetuates confusion concerning the faulty idea but also shows an evident neglect and an *a priori* prejudice. Interestingly, the authors [1] have acknowledged their indebtedness "to the referee for providing thought provoking comments". Presumably these comments played a major role in changing the authors' attitude towards the odd-order ZFS terms in SH from promoting the faulty idea in [2] to an ambiguous presentation in [1]. The paper [1] came to the author's attention during work on a series of extensive review articles [7, 8]. For the benefit of a wider audience, in this Comment we attempt to clarify the confusing points, which led to the rejection of the original manuscript [2] and, which, to a certain extent, still exist in [1].

## 2. Discussion the confusing points

The following points in [1, 2] need clarification as they may lead to further proliferation of the confusion in the electron magnetic resonance (EMR) literature.

**(1) Confusing treatment of the odd-order ZFS terms in SH**

The detailed considerations leading to the conclusion that the admission of the odd-order ZFS terms into SH for the S-state ions in the earlier papers [3, 4] was in fact **incorrect** were presented in [5]. Independently, Grachëv [6] has pointed out additional misunderstandings in the reasoning in [3, 4] and dismissed the admissibility of the odd-order ZFS terms in the generalized SH. Interestingly, the authors made a renewed attempt [9] to justify the odd-order ZFS terms in the SH for the S-state ions on different grounds. On the one hand the authors [9] claim that the relativistic *effective* Hamiltonian for the S-state ions *"replaces the traditional phenomenological spin Hamiltonian"* and admits the odd-order ZFS terms, whereas on the other hand mention that their earlier [3, 4] *"derivation was flawed because it used spin (**S**) rather than total angular momentum (**J**)"*. The paper [9] (quoted as ref. 2 in the original manuscript [2] but omitted in [1]) is another source of confusion on the matter in question. Furthermore, the authors [1] have chosen again, as they did in [9], to ignore the criticism [5, 6] of their earlier attempts [3, 4] to justify the odd-order ZFS terms.

A comparison of the papers [1-4, 9] reveals a specific selection of arguments in various papers by the same authors and thus the meanders of the authors' thoughts on the matter in question. It turns out [5, 10] that the attempts [2-4, 9] are based on serious semantic misunderstandings concerning, e.g., the properties of *"phenomenological SH"* (PSH), as well as the confusion between the crystal field (CF) parameters and the ZFS ones. An example of such misunderstandings may be quoted from [9], where the *"phenomenological spin Hamiltonian for the crystalline electric field"* (sic!) is given as a combination of the terms $A_q^k <r^k> C_q^k$, which in fact represents one form of the *actual* CF Hamiltonian [7, 10]. Keeping in mind the misunderstandings in [9] not much credibility could be given to the new attempt to justify the odd-order ZFS terms presented in the manuscript [2] and, to a certain extent, in [1]. The point is that the odd-order ZFS terms in SH [2, 3, 4, 9] are not allowed by time-reversal invariance as discussed in [5, 6] and partially admitted in [1]. The way, in which the odd-order ZFS terms in SH were introduced in [2, 3, 4, 9], shows that the underlying misunderstandings lead not only to an inappropriate nomenclature, but also to the presence of forbidden terms in SH. This emphasizes strongly the importance of clarification of the underlying semantic issues. For this purpose the meanings of the terms, which are often confused with each other, e.g. *physical* versus *effective* Hamiltonian, *real* versus *effective* versus *fictitious* spin, microscopic SH (MSH), ZFS Hamiltonian, generalized SH (GSH), and phenomenological SH (PSH), have been elucidated in the review [10] and more recently in [7].

It is interesting to compare the corresponding statements in the original manuscript [2] and the published paper [1]. The odd-order ZFS terms are explicitly allowed for certain symmetry cases in Table 1 (identical in [1] and [2]) and Eqs (6, 7a, 7b) in [2] being equivalent to Eqs (9, 9a, 9b), respectively, in [1]. The statement appearing below Eq. (6) in [2] (i.e. Eq. (9) in [1]): *"The odd rank terms are allowed in the 'effective' Hamiltonian representation [2]* (i.e. Ref. 9 here)*"* has been omitted in [1]. The statement in [2]: *"Eqns. (7a)* (i.e. Eq. (9a) in [1]) *and (7b)* (i.e. Eq. (9b) in [1]) *demonstrate that these ZFS Hamiltonians are distinguishable in the relativistic "effective" Hamiltonian representations"* has been replaced in [1] by: *"Equations (9a) and (9b) demonstrate that these two ZFS Hamiltonians cannot be distinguished since only the even rank (k even) terms are allowed by the time reversal symmetry constraint…"*. A sentence has been added at the end of Section 3 in [1]: *"The ZFS odd rank terms, which are not allowed because of the time reversal symmetry constraint in either the conventional phenomenological or generalized spin Hamiltonian formulations have been included to demonstrate that these two symmetries are indistinguishable when these terms are omitted."* The logic used in the latter sentence is quite peculiar. Since the ZFS terms with k odd are not allowed in PSH and GSH [1, 5, 6], these terms, in the first place, should not be included in Table 1 and Eqs (9, 9a, 9b) in [1]. Then no question of the indistinguishability of the two symmetries as well as of the necessity of the omission of these terms would arise. Comparison of the Section 3 in [2] and in [1] indicate a bending of the arguments in [1] for the sake of retaining explicitly the ZFS terms with k odd, since otherwise the whole Section 3 would have been rendered useless in [1]. As pointed out in the review [7], the authors [1] tried to reintroduce the odd-order ZFS terms just using *'tesseral rather that one of the conventional tensor angular momentum operator formulations for the ZFS terms in a PSH'*, while at the same time admitting that these terms *"are not allowed"* on the basis of time-reversal invariance. The latter point has already been discussed in the paper [5] (see also [6]), which, however, received no mention in [1], although it was known to the authors [1, 2] via a private communication (CZR). The convoluted logic and ambiguous presentation of the odd-order ZFS terms in [1] thus adds up to the existing confusion concerning the SH formalisms reviewed in [10, 7].

Notwithstanding this criticism, there is a positive outcome in [1] as compared with [2]. The above direct quotes from the manuscript [2] and the paper [1] (as well as other pertinent statements in [1]) reveal a significant positive change in the authors' attitude towards the odd-order ZFS terms. It appears that in [1] the authors accept, as they stated, e.g. in Abstract in [1], *"... the well known (sic!) conclusion that only even rank zero field splitting terms can be used in the parameterization of the EPR spectra for systems with an odd number of electrons."* **Consequently, it may be hoped that long story of the odd-order ZFS terms in SH [2, 3, 4, 9] will now be put to rest in agreement with the refutations [5, 6].**

For completeness, it is worth mentioning the SH with the *fictitious* spin S'=1/2 for the non-Kramers rare-earth ions with integral J values in the ground state (for details see [7]), e.g. $Tb^{3+}(4f^8)$ ion with the ground multiplet $^7F_6$ (J=6) in ethylsulphates [11, 12, 13]:

$$H = g_\parallel \mu_B B_z S'_z + \Delta S'_x . \tag{1}$$

Eq. (1) includes a ZFS term ($\Delta S'_x$) of **odd** power in the *fictitious* spin operator $S'_x$. Note that, as it follows from the microscopic relation for the parameter $\Delta$ (Ref. 13, p.739), the **admissible** term $\Delta S'_x$ has a different physical origin than the **inadmissible** [5, 6] odd-order ZFS terms in [1, 2, 3, 4, 9], i.e. those with the rank 1, 3, and 5 in Table 1 and Eqs (9) in [1]. Regrettably, the question of the allowed ZFS terms for systems with an 'even number of electrons', i.e. the non-Kramers ions, or systems with integer spin was left open in [1].

**(2) Superfluous tesseral tensor operator notation**

The operators proposed in Eq. (1) in [1, 2] add to the confusing abundance of operator notations existing in the literature as documented in the reviews [10, 8]. The operator notation in [1, 2] is one of the several notations proposed by Buckmaster *et al.* in the last several decades (for a review see [10, 8]), whereas another recent attempt was made by the authors [14]. Since the notation [1, 2] is equivalent, to within a simple numerical factor, to the notations already existing in the literature [15, 16, 17], it is superfluous. Making no effort to relate the 'new' notation with the existing ones [15, 16, 17], the authors [1, 2] only enhance the confusion already widespread in the EMR area [10, 8]. Note that more recently Antonova et al. [18], Ryabov [19], and Tennant et al. [20] have also discussed similar notations. The operator notations [1, 2, 14-20] belong to the group of the tesseral harmonics operator equivalents, which can be subdivided into several subgroups [10] (for a review of more recent literature see [8]). The spin operators defined in [1, 2, 14-18, 20] belong to the category of the **normalized combinations of spherical tensor** (NCST) operators [10, 8] and either are equivalent with each other or differ only by a simple normalization factor [8]. Hence, there is no point in adding yet another operator notation and respective symbols to the existing abundance.

**(3) Inconsistent definition of the Stevens operator notation**

The *"modified Stevens operators"* $O_l^m$ in Eq. (12a) in [1] include explicitly the negative components m<0 and are referred to Abragam and Bleaney [13]. In fact, no negative components of the Stevens operators have been defined in [13], whereas two types of the Stevens operators exist [10, 21]. These are: (i) the ***usual*** (or *conventional*) Stevens operators $O_l^m$ [13], which were originally defined only for $m \geq 0$ [21], and (ii) the ***extended*** Stevens (ES) operators $O_k^q$, which were introduced in [21] as an extension of the former operators and comprise in a unified way also the negative (q < 0) components [10, 8]. No explanation is provided in [1] on how the Stevens operators [13] are *'modified'* and no reference to the extended Stevens operators [21] is made in [1]. More importantly, the general form of the ZFS Hamiltonian in terms of the *"modified Stevens operators"* in Eq. (12a) in [1] includes an inconsistent scaling factor for the second-order ZFS terms, i.e. 1/6, instead of the consistent one prevailing in the literature, i.e. 1/3 [10, 8, 13, 21, 22]. This may be only a misprint, however, as a consequence, the corresponding values in Table 2 in [1] are unreliable since these values may be re-scaled in an inconsistent way thus leading to an incorrect interpretation.

It is worth to make the readers aware of the three serious cases of incorrect relations between the ES ZFS parameters and the conventional ones recently identified in the EMR literature and clarified - for details and references see [23]. They concern: (i) a controversy concerning the second-order rhombic ZFS parameters, which has lead to misinterpretation, in a review article, of several values of either *E* or $b_2^2$ published earlier; (ii) the set of five relations between the extended Stevens ZFS parameters $b_k^q$ and the conventional ones $D_{ij}$ for triclinic symmetry, four of which turn out to be incorrect; and (iii) the omission of the scaling factors $f_k$ for the extended Stevens ZFS parameters $b_k^q$.

**(4) Inadequate discussion of the properties of monoclinic Hamiltonians**

The authors [1, 2] seem to be unaware of the existence of three alternative forms of monoclinic ZFS (CF) Hamiltonians. Importantly, the form of the ZFS (CF) Hamiltonian for monoclinic symmetry depends on the choice of the axis system (x, y, z) with respect to the monoclinic axis $C_2$ for the groups $C_2$ and $C_{2h}$, whereas the monoclinic direction for the group $C_s$ (see, e.g. [24-26]). For monoclinic $C_{2h}$ symmetry the authors [1] use the *'ZFS spin Hamiltonian'* in Eq. (12a) and the *'CEF'* Hamiltonian in Eq. (12b) (note CEF is a standard abbreviation for the 'crystalline electric field', i.e. CF). Both Hamiltonian forms [1] correspond implicitly to a specific choice, namely, the $C_2$ axis parallel to the z-axis. However, no verification is provided in [1, 2] that such choice of the z-axis is valid for the experimental EPR data for $NH_4Ln(SO_4)_2 \cdot 4H_2O$ used in their reinterpretation. The orientation of the monoclinic axis (or direction) in crystal must be first determined from the crystallographic data and hence a proper form of ZFS Hamiltonian must be used in fitting EPR spectra [26]. Without considering the above points the discussion of the monoclinic ZFS and CF Hamiltonians in [1, 2] is inadequate and may lead to misinterpretation of the experimental EPR data.

**(5) Confusion between two physically different Hamiltonians**

The statement *"Eqn. (12a)* (which denotes the ZFS Hamiltonian) *can be expressed in terms of tesseral tensor operators as"* - here Eqn (12b) appears, which denotes the CEF Hamiltonian - indicates a profound confusion of two physically different Hamiltonians in [1, 2]. By changing from one operator notation (the modified Stevens operators in Eq. (12a)) to another (the tesseral tensor operators in Eq. (12b)) the physical nature of a Hamiltonian cannot be changed. The confusion in question is also evident in the last paragraph of Section 4 in [1], where the terms 'CEF' and 'ZFS' are used in various context without a clear distinction between the two physically different quantities. The pertinent examples of this confusion of various degrees occurring in the literature have been discussed in [10, 7].

### 3. Other Points and Summary

In addition to the points discussed above, the following remarks may be made. The reanalysis of the experimental data of others for $Gd^{3+}$ ion in $NH_4Ln(SO_4)_2 \cdot 4H_2O$ in [1] is not physically meaningful since it is based only on changing the notation for the ZFS parameters. No new physical information on the site symmetry can be gained in this way. Moreover, keeping in mind the aspects mentioned in the point 3 and 4 above, the ZFS parameters in Table 2 in [1], especially the second order ones, are unreliable. Note also that the notation for the ZFS parameters used in Table 2 in [1] is rather awkward and it is unlikely others will adopt such notation.

There are no new arguments in [1] to support or disprove the idea of admission of the odd-order ZFS terms into SH, whereas the authors' previous attempts [3, 4, 9] to justify such admission have been sufficiently refuted before in the literature [5, 6]. The literature search indicates only one quotation of [1] by Tennant *et al.* [20], who introduce similar tesseral spherical tensor operators to that defined in [1]. The relationships between various NCST notations will be discussed in details in [8].

Unsubstantiated criticism in [1] of the review [10] calls for a brief comment. The remark below Eq. (12a) in [1]: *'Rudowicz [36]* (Ref. 10 here) *has provided a comprehensive discussion concerning the various notation conventions that have been used for angular momentum operators. However, it should be noted that his discussion contains some errors and misconceptions.'* In the interest of the scientific community, such 'errors and misconceptions', if any, should be explicitly listed and clarified. However, the authors [1] do not do so and hence their criticism is not credible.

*Acknowledgments* My apology is extended to both colleagues whose work has been critically commented on. Note that the major motivation for this paper has been derived from the biblical phrase: *Straighten His Pathways*. The RGC and the City University of Hong Kong supported this work through the research grant: SRG 7001277.